\documentclass[acmsmall,screen,nonacm]{acmart}
\usepackage[shortlabels]{enumitem}
\usepackage{amsthm}

\usepackage{xspace}
\newcommand{\toolname}{ES-Trace\xspace}
\usepackage{lipsum}
\usepackage{booktabs}
\usepackage{balance}
\usepackage{todonotes}
\renewcommand\footnotetextcopyrightpermission[1]{}
\AtBeginDocument{%
  \providecommand\BibTeX{{\normalfont B\kern-0.5em{\scshape i\kern-0.25em b}\kern-0.8em\TeX}}}

\setcopyright{none}
\usepackage[skins]{tcolorbox}
\newtcolorbox{findingbox}{
    enhanced,
    boxrule=0pt,
    frame hidden,
    borderline west={2pt}{0pt}{blue!60!black}, %
    colback=gray!10,                            %
    sharp corners,                             %
    fontupper=\itshape, %
    left=2pt, %
    right=0pt,
    top=1pt,
    bottom=1pt,
    beforeafter skip=1em,
}
\usepackage{colortbl}
\begin{document}

\title{ES-Trace: Auditing Ethical-Sourcing Disclosure of Code Generation Models Beyond Model Cards}
\author{Zhuolin Xu}
\email{zhuolin.xu@mail.concordia.ca}
\affiliation{%
  \institution{Concordia University}
  \city{Montreal}
  \state{Quebec}
  \country{Canada}
}

\author{Haibo Wang}
\email{haibo.wang@mail.concordia.ca}
\affiliation{%
  \institution{Concordia University}
  \city{Montreal}
  \state{Quebec}
  \country{Canada}
}

\author{Shin Hwei Tan}
\email{shinhwei.tan@concordia.ca}
\affiliation{%
  \institution{Concordia University}
  \city{Montreal}
  \state{Quebec}
  \country{Canada}
}

\renewcommand{\shortauthors}{Xu et al.}

\begin{abstract}
Code generation models have been increasingly used in software development, but their development raises ethical-sourcing concerns involving intellectual property, privacy, fairness, labour practices, and environmental impact. Although prior work has defined ethical-sourcing criteria for code generation, it remains unclear how much evidence existing models disclose and where that evidence can be found. We introduce \toolname, a framework for ethical-sourcing disclosure audits that traces disclosed evidence beyond model cards using the Model Documentation Traceability Graph (MDTG), which represents relationships among models, versions, and documentation artifacts. We apply \toolname to 26 models from 10 publishers across 77 documents and 20 ES-CodeGen aspects. Model-card-only auditing yields a mean score of 1.77/5, while resolving the declared references increases it to 2.82/5, with most of the increase arising from documents that the publisher declares in structured metadata. %
The key findings of our study include: (1) social and labour-related aspects remain poorly documented, even when expanding the audit to the full documentation scope, (2) resolving documentation references substantially increases observed disclosure, raising the mean score from 1.77/5 to 2.82/5, (3) model-card-only audits can mischaracterize release-level disclosure changes, and (4) documentation mismatches can associate evidence with the wrong model or version, highlighting the need for explicit model--version binding and consistency across documentation artifacts. Our study calls for reference-aware ethical-sourcing disclosure audits, explicit model--version binding and consistency across documentation artifacts, and stronger documentation of currently underreported social and labour-related aspects. %

\end{abstract}



\maketitle

\section{Introduction}\label{sec:1}

Large language models (LLMs) are widely used to generate, modify, and maintain software. Their development, however, introduces software supply-chain concerns that extend beyond the behavior of the resulting code~\cite{10.1145/3708531,10.1145/3706598.3713787,rahman2025hugginggraph}. Code generation models depend on training data, human contributors, computational infrastructure, and other resources whose acquisition and management can raise questions about intellectual property, contributor rights, privacy, working conditions, and environmental impact~\cite{xu2026makes}. Legal disputes illustrate these concerns. The GitHub Copilot litigation, for example, raises questions about the use of code from public repositories subject to open-source licenses~\cite{copilotlitigation}, while other lawsuits challenge the use of copyrighted works for AI training~\cite{openailawsuit2026}. Environmental impact is another concern: data centers consumed approximately 415~TWh of electricity in 2024, and the International Energy Agency projects demand to reach approximately 945~TWh by 2030, with AI as a major driver~\cite{iea2025energyai}. These concerns have motivated the study of \emph{Ethically Sourced Code Generation} (ES-CodeGen)~\cite{xu2026makes}, which considers whether code generation models are developed through ethical and sustainable practices across their supply chain.

ES-CodeGen was defined through a systematic literature review and a survey of 32 practitioners familiar with code generation models~\cite{xu2026makes}. The study identifies 20 aspects spanning subject rights, equity, accountability, intellectual property rights, code quality, labour rights, environmental sustainability, and other ethical dimensions. However, important questions about the real-world disclosure practices of existing code generation models remain unanswered, particularly whether we can systematically evaluate the evidence they disclose for these aspects. Developers rarely disclose such evidence through the model itself. Instead, they distribute it across model cards, technical papers, dataset documentation, repositories, governance documents, and other development artifacts. We define an \emph{ethical-sourcing disclosure audit} as a systematic examination of publicly disclosed evidence for these aspects and its supporting documentation. Such an audit must identify relevant disclosures and provide an explanation for each assessment. This poses a challenge: \emph{how can an audit systematically identify, attribute, and validate the evidence needed to assess an aspect?}

An \emph{AI Bill of Materials} (AIBOM) specifies an AI system's components, dependencies, provenance, and supply-chain metadata~\cite{safronov2025taibom,nocera2025aiboms}. For code generation models, AIBOMs can link models to training datasets, documentation, and other supply-chain artifacts. Existing generators such as ALOHA~\cite{davino2025aloha} and the OWASP GenAI Security Project AIBOM Generator~\cite{owasp2026aibomgenerator} capture such metadata and references, but do not assess the contents of referenced artifacts. For example, the OWASP generator identifies the technical paper for \path{allenai/OLMo-2-1124-7B-Instruct} but reports energy consumption as missing, although the paper reports 131~MWh of pretraining energy and approximately 52~tCO$_2$eq for OLMo~2~7B~\cite{teamolmo2024olmo2}. Thus, identifying a source does not mean that the evidence it contains has been recovered. %
Similarly, across our 26 models from 10 publishers, the Hugging Face \path{co2_eq_emissions} field is absent from all 26 model cards, although technical papers provide emissions information. Fifteen models reference their own technical report through an \texttt{arxiv:} tag that can be automatically resolved. These cases show that measuring field presence or reference availability can underestimate disclosure when relevant evidence resides in linked artifacts.

In this paper, we address this \emph{evidence-scope problem}: the absence of a disclosure from a model card or generated AIBOM does not imply that it is absent from the documentation reachable from that model. An audit must traverse the documentation relationships exposed by the model repository, retrieve the resulting artifacts, determine the scope of each artifact, and assess whether its contents provide evidence for the audited model and aspect. This requires addressing three challenges. \textbf{First, disclosure is distributed across documentation.} Relevant evidence can appear in model cards, technical papers, dataset cards, code repositories, organization documents, and other artifacts. An audit limited to the model card can therefore miss evidence that is publicly available through declared references. \textbf{Second, references do not establish evidence scope.} Organizations can share documents across model families, variants can inherit documentation from base models, and references can appear in metadata fields or the card text. Resolving a reference is therefore insufficient: the audit must determine which model, variant, and release the target artifact describes before using its contents as evidence. \textbf{Third, documentation evolves independently of model releases.} Developers can add, remove, or revise disclosures without changing the underlying model, while a single document can describe multiple releases or variants. An audit must therefore bind each document to the release it describes before comparing disclosure across releases.

To address these challenges, we propose \toolname, a framework for ethical-sourcing disclosure audits that traces and assesses documented evidence. \toolname builds a \emph{Model Documentation Traceability Graph} (MDTG), which connects publishers, models, and documents, records how each document is referenced and whether it is first- or third-party, and links models through base-model, release, and size relations. 
MDTG preserves the relationships needed to trace evidence across related documentation and to expand the documentation available to an audit beyond the model card. \toolname then applies an \emph{evidence-grounded measurement} %
to assess the evidence available at each scope. 
For each document and ES-CodeGen aspect, an LLM answers five binary evidence questions and provides an evidence snippet for each \emph{Yes}. \toolname derives the aspect score from these answers, and post-hoc evidence checks reject a small set of mechanically identifiable cases in which the snippet does not support the answer.
Separating evidence extraction, score derivation, and validation makes each assessment traceable to its supporting documentation.

Using \toolname, we conduct a systematic study of ethical-sourcing disclosure across 26 code generation models, including 15 code-generation-specific models and 11 general-purpose models that support code generation. The study covers 77 documents and 1,447 document--aspect pairs. %
Our results show that model-card-only evaluation underestimates disclosed ethical-sourcing evidence, and that structured documentation relationships can recover substantial missing evidence. %
The resulting evidence traces also provide a basis for connecting AIBOM components to the documentation and evidence supporting their ethical-sourcing disclosure audit. 

This paper makes the following contributions:
\begin{description}[leftmargin=*,topsep=0pt]

\item[Study:] We present the first systematic study of ethical-sourcing disclosure across 26 code generation models and 77 documents, characterizing where evidence appears, how documentation references expand disclosure scope, and how evidence varies across related models and releases.

\item[Technique for ethical-sourcing disclosure:] We propose \toolname, a reference-aware framework that combines the \emph{Model Documentation Traceability Graph} (MDTG) with an \emph{evidence-grounded measurement} for evaluating ethical-sourcing disclosure across heterogeneous documentation. \toolname uses the MDTG to trace publishers, models, versions, and documentation via reference edges and model relations, and applies a progressive evidence scale of five evidence questions to each document--aspect pair. %
An LLM evaluates increasingly specific evidence requirements and records supporting snippets, while post-hoc checks verify the snippets against the source documents. This produces traceable aspect-level measurements that distinguish disclosure strength from whether the underlying practice satisfies the aspect.

\item[Evaluation:] We apply \toolname to 26 code generation models, 10 publishers, and 77 documents, and show that documentation scope and evidence attribution affect ethical-sourcing assessments. Model-card-only evaluation misses substantial disclosed evidence and can reverse the observed balance of regressions and improvements across releases, while resolved references can associate evidence with the wrong model or version when they are not explicitly bound to their targets. %

\end{description}

In summary, the key findings from our study include:
\begin{description}[leftmargin=*,topsep=0pt]
\item[Social and labour-related aspects remain poorly documented.] At full documentation scope, ``Working Conditions'', ``Fair Wages'', ``Legal Employment'', ``Culture'', and ``Philosophy and Religion'' all score at most 1.62, with 58--85\% of models having no evidence for these aspects.
\item[Resolving references substantially increases observed disclosure.] Expanding beyond model cards raises the mean from 1.77 to 2.82/5, so model-card-only audits understate the evidence in declared documentation. 
\item[Model-card-only audits can mischaracterize release-level disclosure changes.] We treat a lower disclosure score for a newer release as a disclosure \emph{regression} and a higher score as a disclosure \emph{improvement}. Among large release-level changes, model-card-only auditing identifies 14 regressions and 7 improvements, whereas reference resolution identifies 7 regressions and 16 improvements. Model-card-only auditing also misses 19 of the 40 distinct release--aspect changes.

\item[Documentation mismatches can misattribute evidence.] We identify variant, artifact, release, and documentation mismatches that associate evidence with the wrong model or version, highlighting the need for explicit model--version binding and consistency across documentation artifacts.
\end{description}

\section{Related Work}\label{sec:2}

\subsection{Model Documentation and Ethical-Sourcing Criteria}\label{sec:2.1}

Model cards \cite{mitchell2019model} and datasheets for datasets \cite{gebru2021datasheets} establish documentation practices for describing models and datasets, with corresponding templates adopted by platforms such as Hugging Face. Several studies have examined how consistently such documentation is completed. Liang et al. analyze 32,111 model cards and find that environmental impact, limitations, and evaluation are among the least-completed sections \cite{liang2024systematic}. A prior study shows that ethical information is frequently missing from documentation for third-party models and datasets \cite{oreamuno2023state}, while another study finds that documentation for code models on the Hugging Face Hub is thinner than that of AI repositories on GitHub \cite{yang2025ecosystem}. These studies primarily analyze individual model or dataset documents, rather than tracing evidence across related documentation artifacts.%

\noindent\textbf{Aspects in ES-CodeGen.} %
Our audit criteria come from ES-CodeGen \cite{xu2026makes}, which defines 20 aspects based on a literature review and practitioner survey. The taxonomy specifies evidence relevant to ethical sourcing: (1) \textbf{Informed Consent (Opt-in)}, whether data subjects consent to source use; (2) \textbf{Privacy and Security}, protection of private information; (3) \textbf{Diversity}, inclusion of diverse perspectives; (4) \textbf{Fairness}, freedom from bias; (5) \textbf{Representativeness}, coverage of different groups; (6) \textbf{Accessibility}, availability of resources; (7) \textbf{Transparency}, visibility into data and processes; (8) \textbf{Source Acknowledgement}, attribution of sources; (9) \textbf{Licensing}, compliance with licenses; (10) \textbf{Generation Distinctiveness}, distinction between outputs and sources; (11) \textbf{Contamination}, data quality and contamination checks; (12) \textbf{Accuracy}, correctness of generated code; (13) \textbf{Community Development}, benefits to affected communities; (14) \textbf{Philosophy and Religion}, respect for beliefs and sensitivities; (15) \textbf{Culture}, consideration of cultural values; (16) \textbf{Fair Wages}, fair compensation for workers; (17) \textbf{Working Conditions}, safe working conditions; (18) \textbf{Legal Employment}, compliance with labour protections; (19) \textbf{Energy Consumption}, energy use in development; and (20) \textbf{Emissions and Pollution}, environmental emissions and pollution.
ES-CodeGen specifies what evidence is relevant to each aspect but does not prescribe where that evidence should appear. 
Our work connects these ethical-sourcing criteria with the documentation relationships used to locate and attribute such evidence.

\subsection{Ethical and Transparency Auditing}\label{sec:2.2}

Prior work examines ethical and transparency information at different documentation scopes. Gao et al. qualitatively analyze 265 documents from Hugging Face and GitHub and identify six themes of ethical considerations \cite{gao2024documenting}. The ``reference to other materials'' theme appears in 31.70\% of the documents and includes both first- and third-party sources. However, their analysis records the existence and type of a reference without following it to assess the evidence it contains. A large-scale study of 61,665 model cards similarly finds that only 16.8\% contain an ethics section and identifies recurring topics such as responsible use, bias and fairness, and risk and safety \cite{garciabarcelo2026ethical}. These studies characterize ethical information within bounded documentation scopes.

Other approaches assess transparency at the developer or project level. The Foundation Model Transparency Index scores developers using 100 indicators based on public material and developer-submitted reports \cite{bommasani2024foundation}, while openness indices assess whether projects release resources such as code, data, model weights, and scientific documentation \cite{liesenfeld2024rethinking,liesenfeld2023opening}. Such measures provide project-level transparency assessments but do not generally preserve document-level provenance for disclosures or establish whether a document describes the specific model or release under audit.

Several studies extend evidence collection beyond model cards. The AI Transparency Atlas retrieves web materials to score 50 models \cite{mamirov2025transparency}, but the materials are determined by search rather than by references declared in model documentation. Alharbi et al. combine model cards with linked research papers and code repositories and apply a common reproducibility checklist across the resulting scopes \cite{alharbi2026enhancing}. Their setting is particularly close to ours, but the primary paper and repository are selected heuristically, without an explicit representation of how declared references resolve to documents and which model each document describes. Other work uses external resources to generate new model cards \cite{liu2024automatic,zhang2026metagai,zhang2026adaqecg}, with evidence that generated-card quality depends on the availability of supporting resources, particularly associated papers \cite{toma2026automatic}. In contrast, our approach preserves the original documentation, follows the references declared by each model, and checks which model or release each referenced document describes.

\subsection{Software Supply Chain and AIBOM}\label{sec:2.3}

Ethical sourcing is also a software supply-chain concern because model development depends on training data, base models, and the processes used to produce and operate models. Large-scale studies demonstrate gaps in the metadata and licensing information surrounding AI artifacts. Stalnaker et al. examine documentation, supply-chain structure, and licensing across 760,000 models and 175,000 datasets on Hugging Face \cite{stalnaker2026empirical}. Longpre et al. trace more than 1,800 text datasets and find that hosting platforms omit licenses in over 70\% of cases \cite{longpre2024large}, while Jewitt et al. trace 232,270 dataset--model--application chains and find that 62.3\% pass through at least one artifact with no declared license \cite{jewitt2026dont}. These studies establish the importance of declared supply-chain relationships, but primarily analyze metadata and relationship structure rather than evidence contained in referenced documentation.

AIBOM approaches extend the software bill of materials (SBOM) to AI components, provenance, and versioning. A multivocal review identifies traceability as a key benefit while noting immature tooling \cite{nocera2025aiboms}, and SPDX 3.0 introduces an AI profile for representing AI components and relationships \cite{rajbahadur2026building,bennet2024implementing}. Existing AIBOM generators, however, commonly derive inventories from model-card metadata. For example, D'Avino et al. attribute 99.18\% of the data fields missing from their generated AIBOMs to information that is either absent from the model card or present in an unmapped section of it~\cite{davino2025aloha}. Large-scale AIBOM completeness studies similarly rely on the completeness measures that generators provide \cite{erfan2026large}, and model cards, datasheets, and SBOMs rarely provide verifiable evidence of model security \cite{nathanson2025ai}. Such approaches do not determine whether apparently missing information is available in documents referenced by the model card. Software traceability research reconstructs relationships among software artifacts \cite{antoniol2002recovering,antoniol2025recovering}, and studies of links in source-code comments and commit messages show that declared links decay over time \cite{hata20199,xiao202318}. Other work reconstructs AI supply-chain graphs by traversing declared relationships such as \path{base_model} references \cite{jewitt2026permissive,rahman2025hugginggraph}, but focuses on lineage, dependencies, and license propagation.

Our work complements these approaches by connecting declared relationships to document-level evidence. Rather than stopping at a resolved component, we trace a declared reference to its documentation artifact, determine which model or release it describes, and inspect its contents for ethical-sourcing evidence. This links supply-chain traceability with evidence-grounded measurement while preserving provenance from the audited artifact to the supporting disclosure.

\section{Approach}
\label{sec:approach}

\begin{figure}[t]
  \centering
  \includegraphics[width=0.8\textwidth]{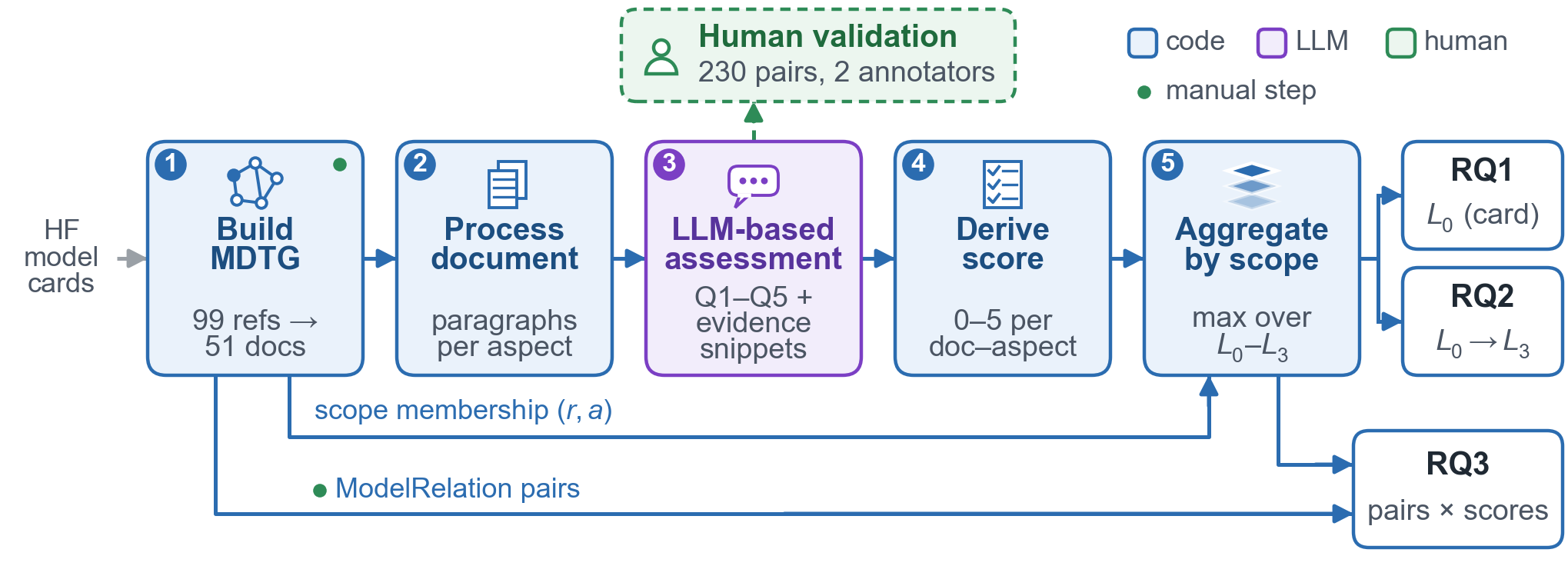}
  \caption{Overview of \toolname. \toolname uses the MDTG to select documents, assign scopes, and pair models for RQ3. The LLM answers Q1--Q5 per document--aspect pair, and \toolname derives and aggregates scores.}

  \Description{A left-to-right five-stage workflow: build the MDTG, process documents, perform LLM-based assessment, derive scores, and aggregate by documentation scope. The MDTG also provides scope membership for aggregation and ModelRelation pairs for RQ3. Human validation is shown above the assessment stage.}
  \label{fig:1}
\end{figure}

\noindent\textbf{Overview of \toolname.}
We measure how much observable evidence about ethical sourcing existing code generation models provide, within and beyond their model cards, using the 20 ES-CodeGen aspects (\S\ref{sec:2}) as measurement dimensions rather than additional scoring criteria. We implement this measurement in \toolname and apply it to a corpus of 26 code generation models from multiple publishers and releases (\S\ref{sec:3.2}). Figure~\ref{fig:1} shows its five stages: \toolname (1) builds the MDTG from Hugging Face (HF) metadata and collects the documents that its $\mathsf{References}$ edges point to (\S\ref{sec:mdtg}), (2) processes documents, (3) performs LLM-based assessments, (4) derives document--aspect scores, and (5) aggregates them by documentation scope (\S\ref{sec:measurement}). The unit of analysis is the \emph{document--aspect pair}. We evaluate the LLM-based assessments on a stratified sample of 230 human-annotated pairs (\S\ref{sec:3.4}).

\subsection{Model Selection and Corpus Construction}
\label{sec:3.2}

\begin{table}[t]
\centering
\caption{The 26 models, one checkpoint per release group (left: code-generation-specific families, right: general-purpose families). \emph{Other sizes} lists sizes of the same release and variant outside the corpus, excluding context-length variants. Bold sizes are the RQ3 same-release baseline (\S\ref{sec:4.3}) and are audited at $L_0$ only.}

\label{tab:models}
\scriptsize
\setlength{\tabcolsep}{2pt}
\resizebox{\textwidth}{!}{%
\begin{tabular}{@{}llrl@{\hspace{5pt}}|@{\hspace{5pt}}llrl@{}}
\toprule
\textbf{Family} & \textbf{Checkpoint} & \textbf{Size} & \textbf{Other sizes} &
\textbf{Family} & \textbf{Checkpoint} & \textbf{Size} & \textbf{Other sizes} \\
\midrule
BigCode      & \texttt{santacoder}                        & 1.1B    & --                       & gpt-oss & \texttt{gpt-oss-20b}                  & 21B$^{\dagger}$ & \textbf{117B}$^{\dagger}$ \\
             & \texttt{starcoder}                         & 15B     & --                       &         & \texttt{gpt-oss-safeguard-20b}        & 21B$^{\dagger}$ & 117B$^{\dagger}$ \\
             & \texttt{starcoder2-15b}                    & 15B     & \textbf{3B, 7B}          & OLMo    & \texttt{OLMo-7B-Instruct}             & 7B   & -- \\
Code Llama   & \texttt{CodeLlama-34b-Instruct-hf}         & 34B     & \textbf{7B, 13B}    &         & \texttt{OLMo-2-1124-7B-Instruct}      & 7B   & \textbf{13B} \\
CodeGemma    & \texttt{codegemma-7b-it}                   & 7B      & --                       &         & \texttt{Olmo-3-7B-Instruct}           & 7B   & -- \\
             & \texttt{codegemma-1.1-7b-it}               & 7B      & --                       &         & \texttt{Olmo-3.1-32B-Instruct}        & 32B  & -- \\
DeepSeek     & \texttt{deepseek-coder-6.7b-instruct}      & 6.7B    & \textbf{1.3B, 33B}       & Phi     & \texttt{phi-1\_5}                     & 1.3B & -- \\
             & \texttt{deepseek-coder-7b-instruct-v1.5}   & 7B      & --                       &         & \texttt{phi-2}                        & 2.7B & -- \\
             & \texttt{DeepSeek-Coder-V2-Instruct}        & 236B$^{\dagger}$ & 16B$^{\dagger}$ &         & \texttt{Phi-3-mini-4k-instruct}       & 3.8B & \textbf{s (7B), m (14B)} \\
Granite      & \texttt{granite-8b-code-instruct-4k}       & 8B      & \textbf{3B, 20B, 34B}    &         & \texttt{Phi-3.5-mini-instruct}        & 3.8B & 42B$^{\dagger}$ \\
             & \texttt{granite-20b-code-instruct-r1.1}    & 20B     & --                       &         & \texttt{Phi-4-mini-instruct}          & 3.8B & -- \\
Kimi         & \texttt{Kimi-K2.7-Code}                    & 1T$^{\dagger}$ & --                &         &                                       &      & \\
Qwen         & \texttt{CodeQwen1.5-7B-Chat}               & 7B      & --                       &         &                                       &      & \\
             & \texttt{Qwen2.5-Coder-7B-Instruct}         & 7B      & \textbf{1.5B} &         &                                       &      & \\
             & \texttt{Qwen3-Coder-30B-A3B-Instruct}      & 30B$^{\dagger}$ & 480B$^{\dagger}$ &         &                                       &      & \\
\bottomrule
\end{tabular}}

\smallskip
\raggedright
$^{\dagger}$~Mixture-of-experts model; total parameters.
\end{table}

\noindent\textbf{Model selection.} %
We constructed a corpus of 26 code generation models from 10 publishers: 15 models explicitly positioned for code generation and 11 general-purpose models from the OLMo, Phi, and gpt-oss families that support code-generation use, all publicly available on HF with a model card. 
We used the HF API through \texttt{huggingface\_hub} to retrieve all models tagged \texttt{code} for text or text-to-text generation (7,221 repositories, retrieved on 2026-06-17) and inspected the 100 most-downloaded among those with recent activity. As this list under-covered some official code-generation families and included many third-party copies, we used it for family discovery only: we retained Qwen, Phi, and BigCode and added eight families that were missing or underrepresented (Code~Llama, DeepSeek, CodeGemma, Granite, Kimi, Nemotron, OLMo, and gpt-oss). %

\noindent\textbf{Releases and checkpoints.}
We then retrieved 1,365 candidate repositories from these families' HF organizations on 2026-08-26. We targeted each family's code generation models and the general-purpose models for OLMo, Phi, and gpt-oss. We grouped repositories into release groups by version patterns in their names and checked these groups against the model cards. We excluded Nemotron because its repositories mix model generations, product tiers, and fine-tuned models from other publishers. Repositories under a separate product name (e.g., OLMoE and Olmo-Hybrid) form their own product lines rather than releases of the family. To avoid over-representing releases with many variants, we manually selected one checkpoint per release group, excluding quantized, converted, and modality-specific repositories. We treated base, instruction-tuned, and specialized checkpoints (e.g., reasoning models) as distinct variants, and checkpoints differing only in context length as the same size. This yielded one to five release groups per family (Table~\ref{tab:models}).

\noindent\textbf{Document collection.}
For each model, \toolname collects five types of documents, following the references described in \S\ref{sec:mdtg}.

\begin{description}[leftmargin=0pt]
\item[(1) Model cards:] document the model, its intended use, limitations, and other model-specific information (e.g., training parameters, used datasets, and evaluation results). %
\item[(2) Technical papers:] describe the model, its development, training, or evaluation.
\item[(3) Dataset cards:] document datasets used to train, fine-tune, or evaluate the model.
\item[(4) Organization cards:] describe the organization responsible for the model development.
\item[(5) GitHub repositories:] describe the model's implementation, development, training/evaluation.
\end{description}

We fetched all documents on 2026-08-27 (UTC) and provide the resulting corpus with SHA-256 checksums to support reproducibility.

\noindent\textbf{Document processing.}
To limit input length, \toolname retrieves, for each technical paper and aspect, paragraphs that match aspect-specific keywords, together with the following paragraph and the section heading. It also retains tables with their captions because quantitative disclosures, such as \texttt{compute} and \texttt{emissions}, may appear in tabular form. Retrieval determines which text the LLM receives but does not determine the audit outcome. %
Since keyword-based retrieval may miss passages that use different terminology, scores for technical papers represent lower bounds on the evidence identified in the retrieved text. \toolname provides the other four document types to the LLM in full.

\subsection{Model Documentation Traceability Graph} %
\label{sec:mdtg}

We propose the \emph{Model Documentation Traceability Graph} (MDTG), which adapts traceability concepts from software engineering~\cite{antoniol2002recovering,antoniol2025recovering} to model documentation. \toolname uses its $\mathsf{References}$ edges to determine which documents are assessed, the attributes of these edges to assign each document to a documentation scope, and its $\mathsf{ModelRelation}$ edges to pair models for RQ3.

We define the MDTG as an attributed directed multigraph
\[
G=(V,E,\tau_V,\tau_E,\alpha),
\]
where $V$ is the set of nodes, $E$ is a set of edge identifiers, $\tau_V:V\rightarrow\mathcal{T}_V$ and $\tau_E:E\rightarrow\mathcal{T}_E$ assign node and edge types, and $\alpha$ assigns attributes to edges. The graph contains three node types: (n1) $\mathsf{Publisher}$, (n2) $\mathsf{Model}$, and (n3) $\mathsf{Document}$, and three edge types: (e1) $\mathsf{Publishes}$, (e2) $\mathsf{References}$, and (e3) $\mathsf{ModelRelation}$.

\noindent\textbf{Publishes.}
For each model $m$, we represent its Hugging Face namespace as a publisher node $p$, connected to $m$ by a $\mathsf{Publishes}$ edge, and denote $m$'s model card by $d_m$.

\noindent\textbf{References.}
A $\mathsf{References}$ edge from model $m$ to document $d$ represents a document explicitly associated with $m$ through its model card or Hub metadata. Each edge records its reference source and the authorship of the target document as $\alpha(e)=(r,a)$,
where $r\in\{\mathsf{metadata},\mathsf{cardText}\}$ identifies whether the reference occurs in a metadata field or in the card text, and $a\in\{\mathsf{firstParty},\mathsf{thirdParty}\}$ 
 denotes document authorship relative to $m$'s publisher. A \emph{first-party document} is authored or maintained by $m$'s publisher, whereas a \emph{third-party document} is authored or maintained by an organization or individual other than $m$'s publisher. 

\toolname automatically extracts metadata references from the Hugging Face Hub, including the \texttt{arxiv:} tag, the \texttt{datasets:} field, and the model namespace, which identifies the publisher and its organization card. This card describes the publisher rather than the model. 
For card-text references, a rule-based classifier identifies candidate document targets in the card, and we manually verify each candidate.
To keep the documentation scope comparable across models, we include at most one publisher-maintained code repository for each model or model family and at most one technical paper for each model. 
For OLMo 3 and OLMo 3.1, whose model cards list only training and evaluation code, we use the OLMo-Core training repository. %
\toolname resolves only existing documents directly referenced by $d_m$ and does not recursively follow their references, so that every assessed document is explicitly associated with $m$. %
For technical papers, we additionally require manual confirmation that the paper describes $m$, another model in the same release, or the release that $m$ updates or builds on (e.g., the OLMo 3 report for OLMo 3.1 and the Qwen3 report for Qwen3-Coder). We exclude method papers that describe only underlying techniques, except for the DeepSeekMoE paper in DeepSeek-Coder-V2's \texttt{arxiv:} tag, which we retain and examine in \S\ref{sec:4.2}.

For each referenced document, we additionally check \emph{scope binding}, which identifies the artifact the document describes, and \emph{version binding}, which identifies the release it describes. These checks detect evidence that describes a related but different model or release (\S\ref{sec:4.2}).

\noindent\textbf{Model Relations.}
A $\mathsf{ModelRelation}$ edge connects two model nodes and is labeled with a relation type from
$
\mathcal{R}_M =
\{\mathsf{BaseModel},\mathsf{NextRelease},\mathsf{SizeVariant}\}.
$
$\mathsf{BaseModel}$ connects a model to the model declared in its \texttt{base\_model:} field and is the only relation explicitly declared by the publisher. $\mathsf{NextRelease}$ connects adjacent releases of the same model family in chronological order. $\mathsf{SizeVariant}$ connects models from the same release that differ in parameter count and is symmetric. We assign $\mathsf{NextRelease}$ and $\mathsf{SizeVariant}$ using release information and repository names (\S\ref{sec:4.3}).

\begin{figure}[htbp]
    \centering
    \includegraphics[width=0.72\textwidth]{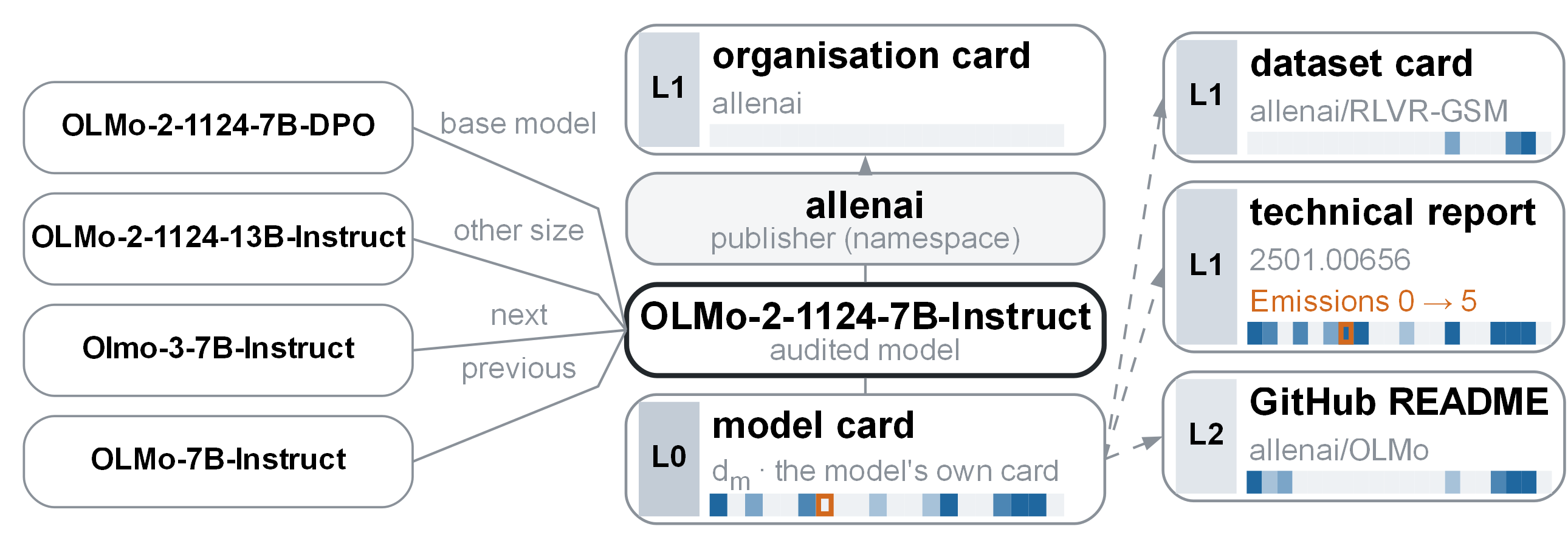} %
    \caption{Example MDTG subgraph for OLMo-2-1124-7B-Instruct. Left: models linked by $\mathsf{ModelRelation}$ edges. Only the base model is declared in its model card. Middle: the publisher, its organization card, the audited model, and its model card $d_m$. Right: documents reached through $\mathsf{References}$ edges (dashed), labeled with their scope. Each strip shows a document's 20 aspect scores (darker means higher). Orange marks Emissions and Pollution, which scores 0 on the model card and 5 in the technical report.}
    \label{fig:2} %
\end{figure}

\noindent\textbf{Documentation scopes.} 
Let $E_m$ denote the $\mathsf{References}$ edges of model $m$, and let $\mathcal{D}(m,L_k)$ denote the set of documents included for $m$ at documentation scope $L_k$. We restrict traversal to one hop from each model: $\mathcal{D}(m,L_k)$ includes the model card and documents directly referenced by $m$, with the scope determining which edge attributes are admitted. %
\[
\begin{aligned}
\mathcal{D}(m,L_0) &= \{d_m\},\\
\mathcal{D}(m,L_1) &= \mathcal{D}(m,L_0) \cup \{d \mid e=(m,d)\in E_m,\ \alpha(e)=(\mathsf{metadata},\mathsf{firstParty})\},\\
\mathcal{D}(m,L_2) &= \mathcal{D}(m,L_1) \cup \{d \mid e=(m,d)\in E_m,\ \alpha(e)=(\mathsf{cardText},\mathsf{firstParty})\},\\
\mathcal{D}(m,L_3) &= \mathcal{D}(m,L_2) \cup \{d \mid e=(m,d)\in E_m,\ \alpha(e)=(r,\mathsf{thirdParty})\text{ for any } r\}.
\end{aligned}
\]
$L_0$ represents model-card-only auditing and approximates the documentation scope supported by existing AIBOM generators, which may record references from model cards without resolving their targets~\cite{davino2025aloha,owasp2026aibomgenerator}. %
We also inspect $\mathsf{BaseModel}$ edges to identify pretraining-data declarations in upstream models (Finding~9).
Figure~\ref{fig:2} shows the MDTG around OLMo-2-1124-7B-Instruct.

\subsection{Evidence-Grounded Measurement}
\label{sec:measurement}

Building on checklist-based approaches to structured LLM evaluation~\cite{lee2025checkeval,cook2024ticking,wei2025rocketeval}, we perform evidence-grounded assessment of model documentation using the 20 ES-CodeGen aspects~\cite{xu2026makes} without modification. For each aspect, the prompt provides a concise aspect description derived from Xu et al. \cite{xu2026makes} and an aspect-specific description of what constitutes evidence for the higher evidence levels. Q1 and Q2 use the same generic disclosure criteria across all aspects, whereas Q3--Q5 use aspect-specific criteria that operationalize increasingly specific and verifiable forms of evidence. The prompt also specifies what is out of scope for the aspect and instructs the LLM to rely only on statements explicitly present in the source.%

\noindent\textbf{Model and settings.} %
We use Gemma 4 31B Instruct (\texttt{gemma-4-31b-it}) \cite{gemmateam2026gemma4} through the Google API with temperature set to 0. We use a single open model because prior work suggests that checklist-based evaluation correlates well with human judgments \cite{wei2025rocketeval} and that mid-sized open models can agree well with human annotators \cite{biskupski2026evaluating}, while evaluations across different LLMs can be inconsistent \cite{nasser2026evaluative}. %
The longest prompt contains 26,198 tokens, well below the model's 256K-token context limit.%
\footnote{Context length as specified in the model card: \url{https://huggingface.co/google/gemma-4-31B-it}.} For the two gpt-oss model cards, the Transparency assessment returned no answer within the output budget. We treat these two pairs as missing rather than imputing them, and discuss the cause and sensitivity in \S\ref{sec:6}.
As LLM-based evaluation performance can vary across tasks \cite{bavaresco2024llms}, we also assess agreement with human annotations (\S\ref{sec:3.4}). To assess repeatability, we rescored the 230 validation pairs twice. All 230 pairs returned a response in the first rerun and 227 in the second, where three requests failed because the service was unavailable. Every returned response reproduced the original score. %

For each document--aspect pair, the prompt provides the name and description of the aspect, and the LLM answers five binary evidence questions (Q3--Q5 shown in generic form):
\begin{description}[leftmargin=*]
\item[Q1:] Does the source mention this aspect at all, in any form?
\item[Q2:] Does the source say anything beyond asserting that the aspect exists or was considered?
\item[Q3:] Does the source specify an object, process, or party, instead of describing one in general terms?
\item[Q4:] Can it be identified by a third party outside the document?
\item[Q5:] Does the source provide a verifiable artifact for it, such as a link, identifier, quotable figure, or named document?
\end{description}
The five questions define a progressive evidence scale, with each successive question requiring more specific or verifiable disclosure. For each document--aspect pair, the LLM records a binary \emph{Yes}/\emph{No} judgment for each question and an evidence snippet from the source for every \emph{Yes} judgment. \toolname assigns the score as the number of consecutive \emph{Yes} judgments from Q1, stopping at the first \emph{No}, yielding an ordinal score from 0 to 5, whose means we report as descriptive summaries. The score measures disclosure strength rather than whether the underlying practice satisfies the aspect. Assessing the underlying practice requires evidence beyond publicly available documentation, such as organizational records or provenance information, and therefore falls outside the scope of this study.

Of the 1,540 document--aspect pairs we attempted to assess, 1,447 received an LLM-based score. 
For 82 paper--aspect pairs, %
aspect-specific retrieval returned no relevant paragraph, and a script-based keyword check of all body sections of the paper found none of the aspect keywords. We assign these pairs a score of 0 for document-type means and exclude the other 11 unscored pairs, in which the aspect terms appear only in bibliography entries or a section heading (9 pairs) or the LLM returned no answer (2 pairs). %
Our analysis of scored pairs uses 1,447 pairs, whereas document-type means that include confirmed paper absences use 1,529 pairs. Since scope scores take the maximum across documents, these zero-valued pairs affect only document-type means, not scope scores. %

\noindent\textbf{Evidence checks.}
As each positive judgment includes an evidence snippet, \toolname automatically applies post-hoc checks. It first verifies that all 2,589 snippets occur in the source text given to the LLM. For the 1,293 snippets supporting Q3--Q5, four further checks flag evidence that does not support the judgment: evidence from usage-oriented sections, evidence inside BibTeX entries, self-referential evidence for ``Source Acknowledgement'', and Q5 evidence without a concrete, verifiable referent. When a check fails, \toolname treats the question as ``No'' and recomputes the score. The checks affect 6 of the 1,447 scored pairs and none of the 230 human-validation pairs (\S\ref{sec:3.4}). Removing them changes scope and document-type means by at most 0.02.

For each model and ES-CodeGen aspect, we define the \textbf{score at documentation scope \(L_k\)} as the maximum document-level score among the documents in \(\mathcal{D}(m,L_k)\). We use the maximum because the audit measures the highest level of evidence available for an aspect within the documentation included in a scope, rather than the consistency of its disclosure across documents. 

\subsection{Manual Annotation to Evaluate the Quality of LLM-Based Assessment}
\label{sec:3.4}

Our goal is to provide independent human judgments to assess the quality of the LLM-based assessment.
As annotating the full population is impractical, we sample 230 document--aspect pairs. We determine the sample size using a standard proportion-based calculation with a 95\% confidence level and a worst-case margin of $\pm5.93\%$ ($p=0.5$), applying a finite-population correction. We stratify the sample \cite{baltes2022sampling} by document type and LLM-based assessment score band and ensure coverage of all aspects. As the allocation is disproportionate across strata, we weight each sampled pair by the inverse of its sampling fraction when estimating population-level agreement.

Two authors of the paper, both PhD students in software engineering with at least one year of experience in performing empirical studies, serve as annotators. Of the 230 sampled pairs, two annotators label 80 shared pairs independently and discuss their disagreements, then divide the remaining 150 pairs between them. They use the same aspect descriptions and five evidence questions as the LLM-based assessment (\S\ref{sec:measurement}) and do not see the LLM outputs.

We report exact agreement and quadratically weighted Cohen's kappa (QWK), which accounts for the ordinal nature of the 0--5 scores and penalizes larger score differences more heavily \cite{cohen1968weighted}.

For the 80 independently annotated pairs, we compare agreement between the two human annotators with agreement between the LLM-based assessment and the human annotations~\cite{ahmed2025can}. Human--human agreement is 0.79 exact agreement (QWK 0.87), while LLM--human agreement is 0.69 exact agreement (QWK 0.80), pooled over the two annotators. The QWK difference between LLM--human and human--human agreement is $-0.08$ (95\% CI $[-0.19, 0.02]$). These results indicate that the LLM-based assessment achieves substantial agreement with human annotation, with no statistically significant difference from inter-annotator agreement. Across all 230 pairs, the design-weighted LLM--human agreement is 0.76 exact agreement (QWK 0.83).

\section{Evaluation Results}
\label{sec:4}
We answer three research questions, from what model cards disclose, to what following their declared references adds, to how disclosure changes across releases and differs across families:
\begin{description}[leftmargin=*]
\item[RQ1:] What do model cards disclose about the 20 ethical-sourcing aspects?
\item[RQ2:] How much additional disclosure becomes observable when declared references are resolved?
\item[RQ3:] How does observed ethical-sourcing disclosure change across releases within a model family, and how do families compare?
\end{description}

\subsection{RQ1: Model Card Disclosure of Ethical-Sourcing Aspects}
\label{sec:4.1}
We score the 26 model cards across all 20 ES-CodeGen aspects at scope $L_0$ and compare aspects by their mean scores and the proportion of pairs receiving a score of 0. The $L_0$ column in Table~\ref{tab:rq1-aspects} shows the mean score of each aspect. %
\begin{table*}[t]
\centering
\caption{Mean ES-CodeGen disclosure scores across 26 models at each documentation scope ($L_0$: model card; $L_1$: + first-party metadata references; $L_2$: + first-party card-text links; $L_3$: + third-party documents), with $\Delta = L_3 - L_0$. The right columns describe the models' $L_3$ scores. The last column is the share of the 10 families with at least one model scoring 5. Transparency excludes the two gpt-oss models, which have no score.}
\label{tab:rq1-aspects}
\small
\setlength{\tabcolsep}{4pt}
\resizebox{0.8\textwidth}{!}{%
\begin{tabular}{@{}lrrrrrcrrr@{}}
\toprule
 & \multicolumn{5}{c}{Mean score} & \multicolumn{4}{c}{Across models at $L_3$} \\
\cmidrule(lr){2-6}\cmidrule(l){7-10}
\textbf{ES-CodeGen aspect} & $L_0$ & $L_1$ & $L_2$ & $L_3$ & $\Delta$ & Range & Zero (\%) & Models at 5 (\%) & Families at 5 (\%) \\
\midrule
Informed Consent (Opt-in)  & 0.50 & 0.77 & 0.92 & 1.08 & +0.58 & 0--5 & 76.92 & 15.38 & 20.00 \\
Privacy and Security       & 0.77 & 2.42 & 2.58 & 2.58 & +1.81 & 0--5 & 30.77 & 34.62 & 50.00 \\
Diversity                  & 2.73 & 3.85 & 4.12 & 4.50 & +1.77 & 0--5 & 3.85 & 69.23 & 80.00 \\
Fairness                   & 1.15 & 2.85 & 3.15 & 3.15 & +2.00 & 0--5 & 19.23 & 38.46 & 50.00 \\
Representativeness         & 3.38 & 4.00 & 4.31 & 4.31 & +0.92 & 0--5 & 3.85 & 69.23 & 70.00 \\
Accessibility              & 4.15 & 4.58 & 4.65 & 4.65 & +0.50 & 2--5 & 0.00 & 84.62 & 100.00 \\
Transparency               & 4.79 & 4.83 & 4.96 & 4.96 & +0.17 & 4--5 & 0.00 & 95.83 & 100.00 \\
Source Acknowledgement     & 4.27 & 4.54 & 4.73 & 4.73 & +0.46 & 2--5 & 0.00 & 80.77 & 90.00 \\
Licensing                  & 3.54 & 3.69 & 3.69 & 3.69 & +0.15 & 3--5 & 0.00 & 26.92 & 30.00 \\
Generation Distinctiveness & 0.81 & 1.46 & 1.46 & 1.46 & +0.65 & 0--5 & 61.54 & 19.23 & 20.00 \\
Contamination              & 0.50 & 2.38 & 2.38 & 2.46 & +1.96 & 0--5 & 42.31 & 23.08 & 30.00 \\
Accuracy                   & 3.65 & 4.77 & 4.88 & 4.88 & +1.23 & 3--5 & 0.00 & 92.31 & 100.00 \\
Community Development      & 1.88 & 2.81 & 3.46 & 3.69 & +1.81 & 0--5 & 7.69 & 46.15 & 70.00 \\
Philosophy and Religion    & \textbf{0.00} & 1.62 & 1.62 & 1.62 & +1.62 & 0--5 & 61.54 & 15.38 & 30.00 \\
Culture                    & 0.31 & 0.62 & 0.62 & 0.92 & +0.62 & 0--5 & 73.08 & 3.85 & 10.00 \\
Fair Wages                 & \textbf{0.00} & 0.65 & 0.65 & 1.04 & +1.04 & 0--5 & 73.08 & 11.54 & 20.00 \\
Working Conditions         & 0.15 & 0.15 & 0.15 & 0.54 & +0.38 & 0--5 & 84.62 & 7.69 & 10.00 \\
Legal Employment           & 0.23 & 0.65 & 0.81 & 0.96 & +0.73 & 0--4 & 57.69 & 0.00 & 0.00 \\
Energy Consumption         & 2.58 & 3.88 & 3.96 & 3.96 & +1.38 & 0--5 & 11.54 & 57.69 & 60.00 \\
Emissions and Pollution    & 0.19 & 1.35 & 1.35 & 1.35 & +1.15 & 0--5 & 73.08 & 26.92 & 40.00 \\
\midrule
\textbf{Overall} &
\textbf{1.77} & \textbf{2.58} & \textbf{2.71} & \textbf{2.82} & \textbf{+1.05} &
-- & \textbf{34.17} & -- & -- \\
\bottomrule
\end{tabular}}
\end{table*}

Across 518 model-aspect pairs, the mean score is 1.77, 
and 55\% of all pairs receive a score of 0. Disclosure is concentrated in a small subset of aspects: ``Transparency'' (4.79), ``Source Acknowledgement'' (4.27), ``Accessibility'' (4.15), ``Accuracy'' (3.65), and ``Licensing'' (3.54) have the highest mean scores. In contrast, several aspects have low mean scores: ``Philosophy and Religion'' and ``Fair Wages'' score 0 on every card, while ``Working Conditions'' (0.15), ``Emissions and Pollution'' (0.19), ``Legal Employment'' (0.23), ``Culture'' (0.31), ``Informed Consent'' (0.50), and ``Contamination'' (0.50) have low scores. Overall, model-card disclosure focuses on five aspects that are related to model and data governance, whereas six aspects, mostly social and labour-related, are largely neglected. 
\begin{findingbox}
\textbf{Finding 1: Ethical-sourcing disclosure in model cards is uneven across aspects.} Across 518 model-aspect pairs, the mean score is 1.77/5, with 55\% receiving no evidence. Five aspects have mean scores above 3.5, whereas six aspects have mean scores below 0.5, including ``Philosophy and Religion'' and ``Fair Wages'', which score 0 for all models.
\end{findingbox}

Disclosure varies substantially across models but is largely shared within families. At model-card scope, per-model mean scores range from 0.75 (Qwen3-Coder) to 2.70 (StarCoder) among models assessed on all 20 aspects, yet models in the same family differ by at most 0.65 points (median 0.38), whereas family means range from 0.87 (gpt-oss) to 2.58 (BigCode). Code-generation-specific and general-purpose models have similar overall disclosure: their per-model means overlap (0.75--2.70 and 0.68--2.50) and have similar medians (1.60 and 1.95).\footnote{Setting the two unscored gpt-oss Transparency pairs to 5 gives a general-purpose range of 0.90--2.50 and the same medians.} Their aspect-level patterns nevertheless differ: general-purpose models score higher on ``Fairness'', ``Community Development'', and ``Contamination'', whereas code-generation-specific models score higher on ``Accuracy'', ``Generation Distinctiveness'', and ``Informed Consent'', the latter two largely because of the three BigCode models. These results complement practitioner perceptions reported in the ES-CodeGen study~\cite{xu2026makes}, where 68.8\% of participants considered none of the existing models to align with ES-CodeGen, while others identified Llama, Qwen, StarCoder, and OLMo 2 as partially or fully aligned. StarCoder also has the highest model-card disclosure score in our corpus. The complete $L_0$ model-by-aspect matrix will be included in our replication package (see Data Availability). %

\begin{findingbox}
\textbf{Finding 2: Model-card disclosure differs mainly between model families, not between code-generation-specific and general-purpose models.} Models in the same family differ by at most 0.65 points, while family means range from 0.87 to 2.58. The two model types have overlapping scores.
\end{findingbox}

\subsection{RQ2: Additional Disclosure from Reference Resolution}
\label{sec:4.2}

Among our five document types, only model cards~\cite{mitchell2019model,hf2026modelcards} and dataset cards~\cite{gebru2021datasheets,hf2026datasetcards} have content standards, and no comparable standard exists for technical papers, GitHub READMEs, or organization cards. These standards address 14 of our 20 aspects, yet technical papers provide the strongest disclosure for ten of these aspects, while cards lead on only four.
Emissions provide the clearest example: Hugging Face documents a \path{co2_eq_emissions} metadata field, but the field is absent from all 26 model cards, and the mean Emissions and Pollution score is 0.19 for model cards compared with 1.88 for technical papers. Overall, model cards and dataset cards also have lower mean scores (1.77 and 1.58, respectively) than technical papers (2.47).

\begin{findingbox}
\textbf{Finding 3: Documentation standards do not determine where disclosure appears.} Although model-card and dataset-card standards address 14 of our 20 aspects, technical papers provide the strongest disclosure for ten of these aspects, while model or dataset cards lead on only four. The \path{co2_eq_emissions} field is absent from all 26 model cards despite emissions evidence appearing in technical papers.
\end{findingbox}

\begin{table}[t]
\centering
\caption{Scores at $L_3$ ($\cdot$: 0, --: no score), with aspects ordered by their $L_0$ mean. Sum$_{L0}$ and Sum$_{L3}$ add up the assessed aspects (at most 100, or 95 for the two gpt-oss models), and Gain $=$ Sum$_{L3} -$ Sum$_{L0}$.}

\label{tab:matrix-l0}
\footnotesize
\setlength{\tabcolsep}{0.9pt}
\setlength{\tabcolsep}{1.6pt}%
\resizebox{0.8\textwidth}{!}{%
\begin{tabular}{@{}llccccccccccccccccccccrrr@{}}
\toprule
Family & Model & \rotatebox{90}{Transp.} & \rotatebox{90}{Src.\ ack.} & \rotatebox{90}{Access.} & \rotatebox{90}{Accuracy} & \rotatebox{90}{Licensing} & \rotatebox{90}{Repr.} & \rotatebox{90}{Diversity} & \rotatebox{90}{Energy} & \rotatebox{90}{Comm.\ dev.} & \rotatebox{90}{Fairness} & \rotatebox{90}{Gen.\ dist.} & \rotatebox{90}{Priv.\ \& sec.} & \rotatebox{90}{Contam.} & \rotatebox{90}{Consent} & \rotatebox{90}{Culture} & \rotatebox{90}{Legal emp.} & \rotatebox{90}{Emissions} & \rotatebox{90}{Work.\ cond.} & \rotatebox{90}{Fair wages} & \rotatebox{90}{Phil.\ \& rel.} & \rotatebox{90}{Sum$_{L0}$} & \rotatebox{90}{Sum$_{L3}$} & \rotatebox{90}{Gain} \\
\midrule
BigCode & SantaCoder & 5 & 5 & 5 & 5 & 5 & 5 & 5 & 5 & 5 & 2 & 5 & 5 & 4 & 5 & $\cdot$ & 1 & $\cdot$ & $\cdot$ & 4 & $\cdot$ & 52 & 71 & +19 \\
 & StarCoder & 5 & 5 & 5 & 5 & 5 & 5 & 5 & 5 & 5 & 5 & 5 & 5 & 5 & 5 & $\cdot$ & 2 & 5 & $\cdot$ & 5 & 5 & 54 & 87 & +33 \\
 & StarCoder2 & 5 & 5 & 5 & 5 & 5 & 5 & 5 & 5 & 5 & 5 & 5 & 5 & 5 & 5 & $\cdot$ & 2 & 5 & $\cdot$ & 3 & 5 & 49 & 85 & +36 \\
\midrule
Code Llama & Code Llama & 5 & 5 & 5 & 5 & 3 & 5 & 5 & 5 & 5 & 5 & 3 & 3 & 4 & $\cdot$ & $\cdot$ & $\cdot$ & 5 & $\cdot$ & 3 & 5 & 43 & 71 & +28 \\
\midrule
CodeGemma & CodeGemma 1.0 & 5 & 5 & 5 & 5 & 4 & 4 & 4 & 4 & 5 & 3 & 3 & 5 & $\cdot$ & $\cdot$ & $\cdot$ & $\cdot$ & $\cdot$ & 2 & $\cdot$ & $\cdot$ & 49 & 54 & +5 \\
 & CodeGemma 1.1 & 5 & 5 & 5 & 5 & 4 & 4 & 4 & 4 & 5 & 3 & 3 & 3 & $\cdot$ & $\cdot$ & $\cdot$ & $\cdot$ & $\cdot$ & $\cdot$ & $\cdot$ & $\cdot$ & 45 & 50 & +5 \\
\midrule
DeepSeek & DeepSeek-Coder & 5 & 4 & 5 & 5 & 3 & 5 & 5 & $\cdot$ & 5 & $\cdot$ & $\cdot$ & 2 & $\cdot$ & $\cdot$ & $\cdot$ & $\cdot$ & $\cdot$ & $\cdot$ & $\cdot$ & 4 & 27 & 43 & +16 \\
 & DeepSeek-Coder v1.5 & 5 & 4 & 5 & 5 & 3 & 5 & 5 & $\cdot$ & 5 & $\cdot$ & $\cdot$ & 2 & $\cdot$ & $\cdot$ & $\cdot$ & $\cdot$ & $\cdot$ & $\cdot$ & $\cdot$ & 4 & 19 & 43 & +24 \\
 & DeepSeek-Coder-V2 & 5 & 5 & 5 & 5 & 3 & 4 & 5 & 4 & $\cdot$ & $\cdot$ & 2 & $\cdot$ & $\cdot$ & $\cdot$ & 4 & $\cdot$ & $\cdot$ & $\cdot$ & $\cdot$ & 4 & 21 & 46 & +25 \\
\midrule
Granite & Granite Code r1.0 & 5 & 5 & 5 & 5 & 5 & 5 & 5 & 4 & 5 & 4 & $\cdot$ & 5 & 5 & 4 & 4 & 4 & 5 & 5 & 5 & $\cdot$ & 32 & 85 & +53 \\
 & Granite Code r1.1 & 5 & 5 & 5 & 5 & 5 & 5 & 5 & 4 & 5 & 4 & $\cdot$ & 5 & 5 & 4 & 4 & 4 & 5 & 5 & 5 & $\cdot$ & 33 & 85 & +52 \\
\midrule
Kimi & Kimi K2.7 Code & 5 & 4 & 5 & 5 & 3 & $\cdot$ & $\cdot$ & $\cdot$ & $\cdot$ & $\cdot$ & $\cdot$ & $\cdot$ & $\cdot$ & $\cdot$ & $\cdot$ & $\cdot$ & $\cdot$ & $\cdot$ & $\cdot$ & $\cdot$ & 22 & 22 & +0 \\
\midrule
Qwen & CodeQwen1.5 & 4 & 5 & 5 & 5 & 3 & 2 & 4 & 2 & 3 & 3 & $\cdot$ & $\cdot$ & 4 & $\cdot$ & 2 & $\cdot$ & $\cdot$ & $\cdot$ & $\cdot$ & 2 & 26 & 44 & +18 \\
 & Qwen2.5-Coder & 5 & 5 & 5 & 5 & 3 & 5 & 3 & 2 & 3 & 2 & $\cdot$ & $\cdot$ & 4 & $\cdot$ & $\cdot$ & $\cdot$ & $\cdot$ & $\cdot$ & $\cdot$ & $\cdot$ & 28 & 42 & +14 \\
 & Qwen3-Coder & 5 & 4 & 5 & 5 & 3 & 5 & 5 & 5 & 3 & $\cdot$ & $\cdot$ & $\cdot$ & $\cdot$ & $\cdot$ & $\cdot$ & $\cdot$ & $\cdot$ & $\cdot$ & $\cdot$ & $\cdot$ & 15 & 40 & +25 \\
\midrule
gpt-oss & gpt-oss$^\dagger$ & -- & 2 & 5 & 5 & 3 & 2 & 5 & 5 & 3 & 5 & $\cdot$ & 3 & $\cdot$ & $\cdot$ & $\cdot$ & $\cdot$ & $\cdot$ & $\cdot$ & $\cdot$ & $\cdot$ & 13 & 38 & +25 \\
 & gpt-oss-safeguard$^\dagger$ & -- & 5 & 5 & 5 & 3 & 2 & 5 & 5 & 5 & 5 & $\cdot$ & 3 & $\cdot$ & $\cdot$ & $\cdot$ & $\cdot$ & $\cdot$ & $\cdot$ & $\cdot$ & $\cdot$ & 20 & 43 & +23 \\
\midrule
OLMo & OLMo & 5 & 5 & 5 & 5 & 5 & 5 & 5 & 5 & 5 & 5 & $\cdot$ & 3 & 4 & 5 & $\cdot$ & 2 & 5 & $\cdot$ & 2 & $\cdot$ & 44 & 71 & +27 \\
 & OLMo 2 & 5 & 5 & 5 & 4 & 5 & 5 & 4 & 5 & 3 & 2 & 2 & $\cdot$ & 4 & $\cdot$ & $\cdot$ & 2 & 5 & $\cdot$ & $\cdot$ & $\cdot$ & 35 & 56 & +21 \\
 & OLMo 3 & 5 & 5 & 5 & 5 & 4 & 5 & 5 & 5 & 4 & 5 & 5 & 5 & 5 & $\cdot$ & $\cdot$ & 2 & $\cdot$ & $\cdot$ & $\cdot$ & 4 & 35 & 69 & +34 \\
 & OLMo 3.1 & 5 & 5 & 5 & 5 & 4 & 5 & 5 & 5 & 4 & 5 & 5 & 5 & 5 & $\cdot$ & $\cdot$ & 2 & $\cdot$ & $\cdot$ & $\cdot$ & 4 & 35 & 69 & +34 \\
\midrule
Phi & Phi-1.5 & 5 & 5 & 5 & 5 & 3 & 5 & 4 & 5 & 3 & 5 & $\cdot$ & $\cdot$ & 4 & $\cdot$ & 2 & $\cdot$ & $\cdot$ & $\cdot$ & $\cdot$ & $\cdot$ & 41 & 51 & +10 \\
 & Phi-2 & 5 & 5 & 3 & 3 & 3 & 5 & 4 & 4 & 3 & 2 & $\cdot$ & $\cdot$ & 2 & $\cdot$ & $\cdot$ & $\cdot$ & $\cdot$ & $\cdot$ & $\cdot$ & $\cdot$ & 39 & 39 & +0 \\
 & Phi-3 & 5 & 5 & 2 & 5 & 3 & 5 & 5 & 5 & 2 & 4 & $\cdot$ & 2 & $\cdot$ & $\cdot$ & $\cdot$ & 2 & $\cdot$ & $\cdot$ & $\cdot$ & $\cdot$ & 45 & 45 & +0 \\
 & Phi-3.5 & 5 & 5 & 3 & 5 & 3 & 5 & 5 & 5 & 3 & 3 & $\cdot$ & 1 & $\cdot$ & $\cdot$ & 5 & 2 & $\cdot$ & 2 & $\cdot$ & $\cdot$ & 50 & 52 & +2 \\
 & Phi-4 & 5 & 5 & 3 & 5 & 3 & 4 & 5 & 5 & 2 & 5 & $\cdot$ & 5 & 4 & $\cdot$ & 3 & $\cdot$ & $\cdot$ & $\cdot$ & $\cdot$ & 5 & 44 & 59 & +15 \\
\midrule
\multicolumn{2}{@{}l}{Mean} & 5.0 & 4.7 & 4.7 & 4.9 & 3.7 & 4.3 & 4.5 & 4.0 & 3.7 & 3.2 & 1.5 & 2.6 & 2.5 & 1.1 & 0.9 & 1.0 & 1.3 & 0.5 & 1.0 & 1.6 & 35.2 & 56.2 & +20.9 \\
\bottomrule
\end{tabular}}

\smallskip
{\raggedright\footnotesize
$^\dagger$~Transparency is unscored for these models, so their sums cover 19 aspects (Section~\ref{sec:6}).\par}

\end{table}

We next examine how expanding the documentation scope through the MDTG (\S\ref{sec:mdtg}) changes disclosure scores. Table~\ref{tab:matrix-l0} reports the $L_3$ scores of all 26 models, together with their summed model-card scores and the difference between the two. We examine the sources of additional disclosure and whether this evidence remains specific to the audited model.
At the model level, the average summed score increases from 35.2 at $L_0$ to 56.2 at $L_3$, corresponding to an average gain of about 21 points (Table~\ref{tab:matrix-l0}). %
Expanding the audit beyond model cards increases the mean score from 1.77 at $L_0$ to 2.58 at $L_1$, 2.71 at $L_2$, and 2.82 at $L_3$ (Table~\ref{tab:rq1-aspects}), and reduces the share of model-aspect pairs without evidence from 55\% to 34\%. Most of the increase occurs when structured first-party references are resolved at $L_1$, while card-text links and third-party documents contribute smaller additional gains. Structured references account for 86\% of the $L_0$-to-$L_2$ increase in the mean score.

\begin{findingbox}
\textbf{Finding 4: Reference resolution substantially increases observed disclosure.} Expanding the documentation scope from model cards to the declared references increases the mean score from 1.77 at $L_0$ to 2.82 at $L_3$, with most of the increase occurring at $L_1$.%
\end{findingbox}

At $L_3$, ``Culture'' (0.92) and ``Philosophy and Religion'' (1.62) also remain low, and 58--85\% of models have no evidence for ``Legal Employment'', ``Fair Wages'', and ``Working Conditions''. The score for the ``Working Conditions'' aspect remains at 0.15 from $L_0$ through $L_2$ and reaches only 0.54 at $L_3$, while ``Fair Wages'' increases from 0 at $L_0$ to 0.65 at $L_1$ and reaches only 1.04 at $L_3$. %
Among first-party publisher documents, the StarCoder paper is the only one that reaches the highest score on Fair Wages, reporting ``an hourly pay rate of \$7.30, taking into consideration different minimum wage rates across countries...'' 
A third-party exception is the NVIDIA HelpSteer dataset card, which reaches 5 on ``Fair Wages'' by naming the standards its annotation vendor, Scale AI, uses to set pay: ``the Anker Methodology, GISC Impact Sourcing Standard, and UN Sustainable Development Goals''. It states that pay is calculated from several factors, including the project, and listed on the vendor's platform, but reports no wage amount. 
HelpSteer is also the only document scoring above 2 on ``Working Conditions'', and no document reaches the highest score on ``Legal Employment''. The increase comes from a few documents: the ``Fair Wages'' increase at $L_1$ is mainly from first-party papers, whereas the ``Working Conditions'' increase appears only at $L_3$, through a third-party dataset card. These examples provide concrete practices that other model developers can reference when documenting fair-wage policies, particularly by specifying the standards and criteria used to determine compensation. Such examples could help support more comparable and reproducible documentation of labour practices across models. %
\begin{findingbox}
\textbf{Finding 5: Social and labour concerns remain poorly documented.} At $L_3$, ``Working Conditions'', ``Fair Wages'', ``Legal Employment'', ``Culture'', and ``Philosophy and Religion'' all score at most 1.62, with 58--85\% of models providing no evidence for the labour-related aspects. Expanding documentation scope recovers limited labour-related evidence. %
\end{findingbox}

Third-party documentation changes the observed standing of the two Granite models. Among the 26 models, IBM's two Granite models are tied for ninth at $L_2$ when the audit uses IBM's first-party documentation, but %
are tied for second at $L_3$ after incorporating the third-party dataset cards they reference. The additional evidence is also concentrated in documents shared across multiple models. Of the 158 model-aspect pairs whose scores increase from $L_0$ to $L_2$, 64 (41\%) receive their $L_2$ score only through documents linked by at least two audited models. These 64 pairs depend on only eight shared documents: three organization cards, two GitHub READMEs, and three technical reports. 
This indicates that a change in one shared document can affect the scores of several models.
\begin{findingbox}
\textbf{Finding 6: Additional evidence is concentrated in a small set of shared documents.} Of the 158 model-aspect pairs that gain evidence from $L_0$ to $L_2$, 64 (41\%) reach their $L_2$ score only through documents linked by multiple audited models. These pairs depend on just eight shared documents, showing that newly observed evidence can be reused across models rather than being model-specific.%
\end{findingbox}

Our manual inspection also identified four patterns related to documentation mismatches that can misattribute evidence: %
\begin{description}[leftmargin=*,topsep=0pt]
\item[P1: Variant mismatch.]%
Evidence reported for one variant can be attributed to another variant from the same release. The StarCoder2 paper reports emissions for StarCoder2-3B and StarCoder2-7B and states that estimates for StarCoder2-15B will be added later. We found no 15B-specific emissions estimate in the paper or the model card. Because the StarCoder2-15B card references this paper, the paper's Emissions and Pollution score of 5 becomes StarCoder2-15B's score at $L_1$, although the reported measurements describe other variants.%
\item[P2: Artifact mismatch.] The GitHub repositories linked from Qwen2.5-Coder and CodeQwen1.5 model cards have been renamed and now describe Qwen3 models. Thus, the currently linked artifacts no longer describe the audited models. Similarly, DeepSeek-Coder-V2's \texttt{arxiv:} tag points to the earlier DeepSeekMoE paper rather than its own technical report. 

\item[P3: Release mismatch.]
The BigCode organization card, which StarCoder reaches through its namespace, documents both StarCoder and StarCoder2. It states that StarCoder was trained on The Stack, which contains only permissively licensed code, whereas StarCoder2 was trained on The Stack v2, which also includes code with no license. The LLM-based assessment of this card cites the Stack v2 description as Licensing evidence, although StarCoder was trained on The Stack.

\item[P4: Documentation inconsistency.] First-party documents can disagree about the same release. The Code Llama 34B model card reports 65.3 tCO$_2$eq for training nine Code Llama models, whereas the technical paper, updated after the 70B release, reports 228.55 tCO$_2$eq for 12 models. The emissions evidence credited to the model therefore depends on which document the audit reads.

\end{description}

Sensitivity analyses show that these patterns affect individual scores but have limited effect on corpus-level means. Removing the same-release variant evidence lowers the $L_2$ mean from 2.71 to 2.70 and the $L_0$-to-$L_2$ gain from 0.95 to 0.94. Treating the DeepSeekMoE paper and the two renamed repositories as unreachable lowers the $L_1$ mean from 2.58 to 2.55 and the $L_2$ mean from 2.71 to 2.69.

\subsection{RQ3: Disclosure Changes Across Releases and Differences Across Families}
\label{sec:4.3}
We evaluate 10 model families: seven code-generation-specific families (BigCode, Code Llama, CodeGemma, DeepSeek, Granite, Kimi, and Qwen) and three general-purpose families (gpt-oss, OLMo, and Phi).
For the eight families with more than one release, we compare every pair of adjacent releases on each aspect at both model-card scope ($L_0$) and full declared-reference scope ($L_3$). This yields 16 adjacent-release pairs and 638 release-level comparisons. For gpt-oss, the only later release on HF is gpt-oss-safeguard-20b, a safety-specialized build of gpt-oss-20b rather than a new generation. %

As a baseline, we use \emph{same-release size variants}, referring to models from the same release with different parameter sizes. %
For each family, we take the earliest release in which the audited model is available in several sizes, and include all of these sizes (all instruction-tuned or all base). All families qualify except CodeGemma (whose second size is released only as a pretrained model) and Kimi (a single size). The eight groups yield 21 pairs (e.g., the DeepSeek-Coder group contains the 1.3B, 6.7B, and 33B models). Since these pairs share a release, they provide a baseline for model-card score variation without a release transition. We assess the size variants only at model-card scope ($L_0$): their cards are not part of the 77-document corpus, and we do not collect their declared references. We define a \emph{large-change event} as a change of three or more score points on an aspect.

Across all 28 pairs of releases within the same family, the magnitude of score change is more strongly associated with differences in model-card text than with the number of releases separating the models. We measure \emph{score change} as the mean absolute difference across the 20 aspects. Controlling for release distance, textual similarity has a partial correlation of $r=-0.45$ with score change, whereas controlling for textual similarity, release distance has a partial correlation of only $r=0.15$. %
Considering both scopes, large-change events occur in 6.9\% of adjacent-release comparisons, compared with 1.4\% of same-release size comparisons at $L_0$ (ratio $=4.82$, 95\% CI [1.85, 18.62], cluster permutation $p=0.010$). At $L_0$ alone, the rate is 6.6\% (ratio $=4.60$, 95\% CI [1.66, 19.03]). Excluding the gpt-oss pair gives 7.0\% versus 1.4\% (ratio 4.89, $p = 0.015$).
\begin{findingbox}
\textbf{Finding 7: Disclosure changes across releases follow model-card rewriting, not release distance.} Large disclosure changes occur in 6.9\% of adjacent-release comparisons, 4.82 times the rate between same-release size variants, yet release distance is only weakly associated with disclosure change ($r=0.15$). In contrast, model-card text similarity is more strongly associated ($r=-0.45$).
\end{findingbox}

Across the 44 large-change events (21 at $L_0$ and 23 at $L_3$, corresponding to 40 distinct release--aspect changes), we compare disclosure scores between an older and a newer release of the same model family. We treat a lower disclosure score for the newer release as a disclosure \emph{regression} and a higher score as a disclosure \emph{improvement}. At $L_0$, model-card-only auditing identifies 14 regressions and 7 improvements. After resolving the declared references, these counts change to 7 regressions and 16 improvements. Thus, documentation scope can change not only the magnitude but also the observed balance of regressions and improvements across releases.

Model-card-only auditing misses 19 of the 40 distinct release--aspect changes, including 13 improvements and 6 regressions. All 13 missed improvements rely on technical papers: ten on the newer model's technical report, one on the general Qwen3 report cited by Qwen3-Coder-30B, and two on the DeepSeekMoE method paper cited by DeepSeek-Coder-V2, which does not describe that model. Excluding these two attribution errors still leaves 14 improvements versus 7 regressions under reference-aware auditing. These results show that model-card-only auditing can mischaracterize whether disclosure improves or regresses across releases.

OLMo provides a concrete example. None of its four model cards provides emissions evidence, so card-only auditing observes no change across releases. After resolving the technical reports referenced by the model cards, OLMo 1 and OLMo 2 receive an emissions score of 5, whereas OLMo 3 and OLMo 3.1 receive 0. The reports quantify emissions for OLMo 1 and OLMo 2 as ``69.78 tCO$_2$eq'' and ``about 154 tCO$_2$eq'', respectively, while the OLMo 3 report provides no emissions information. Thus, card-only auditing misses a release-level \emph{regression} in emissions disclosure that becomes visible only after resolving the referenced technical reports.

\begin{findingbox}
\textbf{Finding 8: Model-card-only audits can mischaracterize disclosure changes across releases.} Across 44 large-change events, card-only auditing identifies 14 regressions and 7 improvements, whereas reference resolution identifies 7 regressions and 16 improvements. Card-only auditing misses 19 of 40 distinct release--aspect changes, showing that release comparisons should resolve the documentation associated with each model release.
\end{findingbox}

Nine of the 26 models declare datasets in the \texttt{datasets:} field, while 17 declare none. Three of these nine models are BigCode pretrained models that declare their pretraining corpora. Among the six instruction-tuned models that declare datasets, five list only post-training datasets, and only OLMo-7B-Instruct also declares its pretraining corpus. %
An additional inspection shows that following \texttt{base\_model:} links for one to three steps reaches a model that declares pretraining data in four of these five cases. Granite 20B is the exception, as its \texttt{base\_model:} points to the model itself. These upstream lineage links are outside our documentation scope, which follows only references directly declared by the audited model.

OLMo shows how direct reference resolution can miss evidence about pretraining data documented for an upstream model. OLMo 1 directly declares Dolma, whose dataset card provides evidence for \emph{Informed Consent} (score 5). In contrast, OLMo 2, OLMo 3, and OLMo 3.1 directly declare post-training datasets, and none of the evidence in their audited documentation satisfies the \emph{Informed Consent} criteria, resulting in a score of 0 at every documentation scope. Their pretraining data is documented further along the model lineage rather than in their direct dataset declarations. %
\begin{findingbox}
\textbf{Finding 9: Direct reference resolution can miss evidence about pretraining data.} Among six instruction-tuned models that declare datasets, five list only post-training datasets. Following \texttt{base\_model:} links identifies pretraining-data declarations for four of these five models, showing that evidence about pretraining data can reside in upstream models rather than in documents directly referenced by the audited model.
\end{findingbox}

\begin{table}[t]
\centering
\caption{Mean score of each family at three documentation scopes, with the rank among the 10 families in parentheses. \emph{Models} is the number of audited models in the family.}
\label{tab:family_rank}
\footnotesize
\setlength{\tabcolsep}{3pt}
\begin{tabular}{@{}lcccccccccc@{}}
\toprule
\emph{Family} & BigCode & CodeGemma & Phi & Code Llama & OLMo & Granite & Qwen & DeepSeek & Kimi & gpt-oss \\
\emph{Models} & 3 & 2 & 5 & 1 & 4 & 2 & 3 & 3 & 1 & 2 \\
\midrule
$L_0$ & 2.58 (1) & 2.35 (2) & 2.19 (3) & 2.15 (4) & 1.86 (5) & 1.63 (6) & 1.15 (7) & 1.12 (8) & 1.10 (9) & 0.87 (10) \\
$L_2$ & 4.05 (1) & 2.60 (5) & 2.46 (6) & 3.55 (2) & 3.31 (3) & 2.90 (4) & 2.10 (9) & 2.20 (7) & 1.10 (10) & 2.13 (8) \\
$L_3$ & 4.05 (2) & 2.60 (5) & 2.46 (6) & 3.55 (3) & 3.31 (4) & 4.25 (1) & 2.10 (9) & 2.20 (7) & 1.10 (10) & 2.13 (8) \\
\bottomrule
\end{tabular}
\end{table}

To examine whether family comparisons depend on documentation scope, we compute the average of the scores of each family's models and rank the 10 families separately at $L_0$, $L_2$, and $L_3$ (Table~\ref{tab:family_rank}). %
BigCode remains among the two highest-scoring families at all three scopes. In contrast, CodeGemma and Phi rank second and third at $L_0$ but fall to fifth and sixth at $L_2$, because their scores increase by only 0.25 and 0.27, respectively, whereas every other family except Kimi gains at least 0.95. Granite shows the largest change: its rank improves from sixth at $L_0$ to fourth at $L_2$ and first at $L_3$, where its gain comes from third-party dataset cards.

This scope dependence is also reflected in a comparison with an independent transparency assessment. The 2024 Foundation Model Transparency Index (FMTI)~\cite{bommasani2024foundation} scores developers rather than individual models. Six developers represented in our corpus are also covered by FMTI: BigCode (SantaCoder, StarCoder, StarCoder2), Meta (Code Llama), IBM (Granite Code), Microsoft (Phi), OpenAI (gpt-oss), and Google (CodeGemma). As each developer contributes exactly one family to our corpus, we pair each developer's FMTI score with the corresponding family-level score. Across these six developers, the Spearman rank correlation increases from $\rho = 0.26$ at $L_0$ to $0.60$ at $L_2$ and $0.66$ at $L_3$. 
This comparison is not a direct validation: FMTI scores developers on 100 indicators for flagship models, most of which are not in our corpus, and incorporates developer-submitted reports, whereas part of Granite's $L_3$ gain comes from third-party dataset cards.

\begin{findingbox}
\textbf{Finding 10: Reference resolution changes family-level disclosure patterns.} Family rankings change substantially as documentation scope expands, with the Granite family moving from sixth at $L_0$ to first at $L_3$. Across six developers overlapping with FMTI, the rank correlation with our family scores increases from $\rho=0.26$ at $L_0$ to $0.66$ at $L_3$.
\end{findingbox}

\section{Implications}\label{sec:5}
We discuss several implications for the model developers and researchers:

\noindent\textbf{Implication 1: Ethics-related audits should trace documentation beyond model cards.}
Our study focuses on ethical-sourcing disclosure audits, but the documentation-scope problem can also affect other audits of ethical properties. Model cards provide an incomplete view of ethics-related disclosure. Resolving their declared references substantially increases the observed disclosure, with the mean score increasing from 1.77 at $L_0$ to 2.82 at $L_3$ (Finding 4). %
Prior work~\cite{garciabarcelo2026ethical} analyzed more than 60,000 model cards and found that only 16.8\% contain an ethics section. Such measurements capture disclosure available in model cards, but can underestimate disclosure when models link to additional documentation that contains relevant evidence. Ethics-related audits should therefore treat model cards as entry points and trace their declared documentation rather than using them as complete audit records.

\noindent\textbf{Implication 2: AIBOM measurements should distinguish field absence from documentation absence.}
An empty AIBOM field does not establish that the corresponding evidence is absent. The \path{co2_eq_emissions} field is absent from all 26 model cards, while technical papers include more environmental disclosure (Finding 3). Our study also reveals that resolving model-card references recovers evidence for aspects that appear poorly documented at $L_0$ (Table~\ref{tab:rq1-aspects}). %
AIBOM-based measurements should therefore report the documentation scope and distinguish missing fields from missing evidence in the reachable documentation.

\noindent\textbf{Implication 3: AIBOM references should bind evidence to model and version.}
Resolving a reference does not establish evidence relevance. We identify variant mismatch, artifact mismatch, release mismatch, and documentation inconsistency across artifacts (\S\ref{sec:4.2}). Shared documents also contribute evidence to multiple audited models, further complicating model-specific attribution (Finding 6). AIBOMs should represent references with their source, authorship, scope binding, and version binding, rather than treating a URL as sufficient provenance, and should record \texttt{base\_model:} lineage so that pretraining-data evidence documented upstream remains reachable (Finding 9).

\noindent\textbf{Implication 4: Documentation templates should explicitly support underreported ethical-sourcing aspects.}
Expanding the documentation scope recovers existing evidence, but it does not resolve aspects that developers rarely document. Labour-related aspects remain poorly disclosed even after tracing first-party and third-party documentation (Finding 5), while the \path{co2_eq_emissions} field is absent from all 26 model cards (Finding 3). Model documentation templates should therefore provide explicit, structured support for underreported aspects such as labour practices and environmental impacts, rather than relying on general-purpose documentation fields.

\noindent\textbf{Implication 5: Release and family comparisons require reference-aware measurement.}
The measured disclosure of a model depends on the documentation scope used by the audit. Analysis that relies only on model cards can miss disclosure changes that become identifiable after resolving references, including improvements supported by technical papers (Finding 8). Reference resolution can also change the relative disclosure levels of model families (Finding 10). Researchers should therefore report the documentation scope used for release and family comparisons and avoid treating model-card-based analysis as a complete measure of model documentation.

\section{Threats to Validity}\label{sec:6}

\noindent\textbf{External.} Our findings may not generalize beyond the 26 publicly available models in our corpus. We mitigate this by selecting models from 10 families across 10 publishers, but our percentages describe this corpus rather than all models on Hugging Face. As the corpus contains only official releases, it may over-represent better-documented models. Documentation can also change after our 2026-08-27 snapshot.%

\noindent\textbf{Internal.} We assess agreement with the LLM using human annotations (\S\ref{sec:3.4}) \cite{halpin2024intercoder,kilicoglu2021toward}. \toolname's post-hoc evidence checks verify that the LLM's evidence snippets occur in the cited documents and flag unsupported Q3--Q5 evidence. We use a Gemma model as the sole LLM, so scores may differ under other LLMs. As the corpus includes two CodeGemma models, the LLM may also favour its own family. Only six validation pairs come from these models' cards, too few to estimate family-specific bias. For gpt-oss-20b and gpt-oss-safeguard-20b, the LLM returned no Transparency score because its reasoning exhausted the output budget. This failure persisted across independent calls. We treat these pairs as missing rather than imputing scores. Assigning 0 or 5 bounds the Transparency aspect means at $L_0$ and $L_3$ to 4.42--4.81 and 4.88--4.96, respectively. We use clustered resampling and permutation tests to account for dependence among pairs from the same documents and model versions. We also assess sensitivity to mismatched variant evidence, renamed or misattributed references, and post-hoc evidence checks (\S\ref{sec:measurement}, \S\ref{sec:4.2}). Our card-text classifier missed CodeGemma's technical report, so CodeGemma's $L_2$ and $L_3$ scores omit it.

\noindent\textbf{Construct.} Our scores measure the strongest evidence disclosed within the one-hop documentation scope, not whether the underlying practice satisfies an ES-CodeGen aspect (\S\ref{sec:measurement}). We mitigate this threat by defining the construct as \emph{ethical-sourcing disclosure}, using aspect-specific evidence criteria, and requiring supporting snippets for positive judgments. We therefore interpret scores as disclosure strength rather than direct assessments of ethical practice. As the human annotators use the same evidence questions as the LLM, the validation tests how consistently these questions are applied, not the questions themselves.

 \section{Conclusion}\label{sec:conclusion}
Our study examines ethical-sourcing disclosure across 26 code generation models, 10 publishers, and 77 documents. We introduce \toolname, which traces documentation relationships with the Model Documentation Traceability Graph (MDTG) and assesses 20 ES-CodeGen aspects using an evidence-grounded measurement. Our results show that model-card-only audits provide an incomplete view of disclosure: referenced documentation adds evidence that model cards lack, such as emissions figures, while social and labour-related aspects remain poorly documented. %
Documentation scope also changes some comparisons across releases and model families, and documentation mismatches can associate evidence with the wrong model or version. These findings motivate reference-aware audits that report their documentation scope and verify evidence against the model and version it describes. We call for AIBOMs that distinguish missing fields from missing evidence and explicitly represent model and version relationships, together with documentation practices that improve coverage of underreported ethical-sourcing aspects.

\section*{Data Availability}
\label{sec:data}
ES-Trace with its prompt, the assessed documents with their SHA-256 checksums, the MDTG relations, all LLM-based assessments with their evidence snippets, the human annotations and validation sample, and the scripts that reproduce the results in this paper will be made publicly available upon publication.

\bibliographystyle{ACM-Reference-Format}
\bibliography{refs}

\end{document}